\documentstyle[12pt,preprint,prl,aps]{revtex}
\begin{document}
\tighten
\title{Truncation method for Green's functions in time-dependent fields} 
\author{Tobias Brandes 
}
\address{Institute of Physics, University of Tokyo, 
Komaba 3-8-1, Tokyo 153, Japan}
\draft
\date{\today}
\maketitle
\date{\today}
\begin{abstract}
We investigate the influence of a time dependent, homogeneous electric field 
on scattering properties of non-interacting electrons 
in an arbitrary static potential. We develop a method to calculate the (Keldysh) 
Green's function in two complementary approaches.
Starting from a plane wave basis, a
formally exact solution is given in terms of the 
inverse of a matrix containing infinitely many 'photoblocks' which can be 
evaluated approximately by truncation. In the exact eigenstate basis of the 
scattering potential, we obtain a version of the Floquet state theory in the 
Green's functions language. The formalism is checked for cases  such as a 
simple model of a double barrier in a strong electric field. Furthermore, an 
exact relation between the inelastic scattering rate due to the microwave and 
the AC conductivity of the system is derived which in particular holds near or 
at a metal--insulator transition in disordered systems. 
\end{abstract} 
\section{Introduction}
The influence of a time-dependent electric field on properties of a quantum 
mechanical system is the topic of active recent research. An essential point 
in these investigations is that the time-dependent field is not from the 
beginning treated in perturbation theory (e.g. in linear response), but is 
rather considered as inherent part of the system itself. By this, one has to 
deal with conditions of a non-equilibrium situation under which the quantity 
of interest, e.g. a tunnel current or the response to an additional DC field, 
has to be determined. There is a considerable variety of physical systems 
studied so far. Many works are related to tunneling, such as driven quartic 
double wells \cite{Hae95}, driven quantum wells \cite{Wag96} or tunneling 
through laser irradiated wells \cite{AIP96}. In interacting systems, 
investigations concentrated on tunneling through quantum dots in time-varying 
fields \cite{Kouetal94,BS94}. 

Another class of investigations are related to transport experiments as, 
e.g., in the quantum Hall regime \cite{Meietal92,Sunetal94,Meietal96}, where 
a microwave gives rise to peculiar changes of the DC conductivities. 
Calculations have been done on photo-induced transport through quantum point 
contacts \cite{FH93}, or on microwave induced changes of the DC conductivity 
of a disordered system \cite{Bra96}. Recently, Holthaus and Hone \cite{HH96} 
studied the Wannier-Stark ladder in an ac-driven one-dimensional tight-
binding system, where in 
presence of disorder-induced localization the localization length is 
controlled by the time-dependent field. 

In spite of this great variety of investigations, the main theoretical focus 
so far has been on the zero- or (quasi) one-dimensional case. It is therefore 
desirable to develop a calculation scheme that is applicable in higher
dimensions and for arbitrary static potentials, in particular in view of 
applications as the quantum Hall effect ($d=2$) or the Anderson transition
($d=3$). In this paper, we set up a general formalism to calculate the Green's 
function of non--interacting electrons moving in an arbitrary static 
potential under the influence of a time--dependent electric field with 
frequency $\Omega $.                      
We use the Dyson equation to obtain the Keldysh Green's function in two different 
calculation schemes. First, we regard the static potential as 
a perturbation and include the electric field in the unperturbed Hamiltonian.
We give a formally exact solution  which  can be evaluated approximately by 
inverting a truncated matrix containing a finite number of 'photo-blocks'. 
The advantage of this method is its exactness in the electric field; it 
furthermore sums up the static potential to infinite order and is 
perturbative in the higher {\em Fourier components} of the Green's function  
which correspond to the 'center--of--mass' time coordinate. Numerical 
examples of this 'truncation method' are presented for the trivial case of a 
constant potential, which allows comparison to an exact result, and a simple 
model for the transmission through a double barrier structure in a strong 
electric field. There, our method qualitatively reproduces previous results 
by other authors \cite{IPT94,AIP96}. In particular, the examples show that 
already a small number of 'photo-blocks' is sufficient to obtain convergence. 

Second, starting from the exact eigenstates of the static potential, again an 
exact formal solution is derived in which the Green's function is represented 
as the inverse of an infinite tridiagonal matrix, which is the Green's 
function analogue of the Floquet state Hamiltonian. This approach in 
particular is useful in situations where the static scattering problem
is already solved (say by numerical diagonalization) and one is interested in 
the effect of an additional, time-dependent electric field.
We calculate the electron self energy of which the diagonal part yields an 
inelastic scattering time $\tau_{MW}$  due to the influence of the electric 
field. In the limit of $\hbar\Omega \ll E_{F}$ (where $E_{F}$ is the Fermi 
level of the system without electric field) we obtain an exact relation 
between $\tau_{MW}$ and the zero temperature AC-conductivity of the system 
without electric field. 
We calculate  $\tau_{MW}$ for microwave radiation in the integer quantum Hall 
effect. From the fact that $\tau_{MW}$ alone is insufficient to interpret 
experimental data, we conclude that one rather has to investigate the full 
Green's function in presence of the time--dependent electric field. 

The paper is organized as follows. In section \ref{generalformulation},
we set up the general formalism (Dyson's equation), in section \ref{ip}
we specify to the plane wave basis and derive the exact expression
for the Green's function which is evaluated numerically.
In section \ref{fieldperturbation}, we give a formulation in the eigenstate 
basis of the static potential and discuss the electric field induced
scattering rate, which again is derived from the formally exact result. 
A short conclusion is given in section \ref{conclusion}.

\section{General formulation}
\label{generalformulation}
\subsection{Dyson equation}
The starting point in each of our approaches is the Dyson equation for the 
Keldysh Green's function matrices. We start from a basis of eigenstates 
labeled with $\alpha $ and define Green's functions \cite{Mahan} 
$iG_{\alpha \beta }^{T}(t_{1},t_{2}):=\langle Tc_{\alpha }(t_{1})c_{\beta 
}^{\dagger}(t_{2})\rangle$,           
$iG_{\alpha \beta }^{\tilde{T}}(t_{1},t_{2}):=\langle \tilde{T}c_{\alpha 
}(t_{1})c_{\beta }^{\dagger}(t_{2})\rangle$, 
$iG_{\alpha \beta }^>(t_{1},t_{2}):=\langle c_{\alpha 
}(t_{1})c_{\beta }^{\dagger}(t_{2})\rangle$, 
$iG_{\alpha \beta }^<(t_{1},t_{2}):=-\langle c_{\beta 
}^{\dagger}(t_{2})c_{\alpha }(t_{1})\rangle$, 
where $T$ ($\tilde{T}$) denote (anti)-chronological time ordering.  Furthermore, the brackets 
$\langle\rangle$ denote as usual an average over a quantum state of the 
system (not necessarily the stationary state). 
The Green's functions are written in a matrix block 
\begin{eqnarray}
\label{matrix}
G(t_{1},t_{2})=
\left( \begin{array}{cc}
G^{T}(t_{1},t_{2}) & -G^{<}(t_{1},t_{2})  \\ 
G^{>}(t_{1},t_{2})   &
-G^{\tilde{T}}(t_{1},t_{2})  
\end{array}\right), 
\end{eqnarray}
for which the Dyson-integral equation has the same form as in equilibrium 
theory,
\begin{eqnarray}
\label{dyson}
G(t_{1},t_{2})=G^{0}(t_{1},t_{2})+\int\int dtdt'G^{0}(t_{1},t)\Sigma (t,t')
G(t',t_{2}),
\end{eqnarray}
where the matrix $\Sigma$ is composed of $\Sigma^{T}$, 
$-\Sigma^{\tilde{T}}$, $\Sigma^{>}$, $-\Sigma^{<}$, in analogy to 
Eq.~(\ref{matrix}). Here, $G^{0}$ denotes the unperturbed Greens' function, 
e.g. $iG_{\alpha \beta }^{T,0}(t_{1},t_{2}):=\langle T \hat{c}_{\alpha 
}(t_{1})\hat{c}_{\beta }^{\dagger}(t_{2})\rangle$, where the electron 
creation (annihilation) operators $\hat{c}^{(\dagger)}_{\alpha }(t)$ are 
given in the interaction picture which will be defined according to the 
splitting of the total time dependent Hamiltonian $H(t)$. This splitting 
defines the way the perturbation theory is performed and will be explained 
below. 

Since we are interested in a Hamiltonian a part of which oscillates with
frequency $\Omega $, it is useful to perform a Fourier analysis according to 
time 'center of mass' and relative coordinates
(''Wigner--coordinates''), namely $T=(t_{1}+t_{2})/2$ 
and $t=t_{1}-t_{2}$. This decomposition is defined according to
\begin{eqnarray}
\label{fourier}
G(t_{1}=T+t/2,t_{2}=T-t/2)=\frac{1}{2\pi}\sum_{N}\int_{-\infty}^{\infty}
d\omega 
e^{-i\omega t}e^{i\Omega N T} G(\omega ,N)
\end{eqnarray}
and correspondingly for $G^{0}$ and $\Sigma $. The inverse transformation is
\begin{eqnarray}
G(\omega ,N)=\int_{-\infty}^{\infty}dt e^{i\omega t}
\int_{0}^{2\pi} \frac{d(\Omega T)}{2\pi}
e^{-i\Omega NT}
G(t_{1}=T+t/2,t_{2}=T-t/2).
\end{eqnarray}
Of special interest is the component $N=0$ which determines the 
average over the 'center-of-mass' coordinate $T$.
In particular, in the case of an equilibrium situation
(no electric field), all components $G(\omega ,N)$ with $N\ne 0$ vanish because
the Green's function depends on the relative coordinate $t=t_{1}-t_{2}$ only.

Inserting Eq.~(\ref{fourier}) into the Dyson equation Eq.~(\ref{dyson}), a 
straightforward calculation yields
\begin{eqnarray}
\label{dyson2}
G(\omega ,N)&=&G^{0}(\omega ,N)+\sum_{N_{1} N_{2}}G^{0}\left(\omega 
+\frac{N_{1}-N}{2}\Omega ,N_{1}\right)\Sigma \left(\omega +
\frac{N_{1}-N_{2}}{2}\Omega ,N-N_{1}-N_{2}\right)\nonumber\\
&\times&
G\left(\omega 
+\frac{N-N_{2}}{2}\Omega ,N_{2}\right).
\end{eqnarray}
In this paper, we concentrate on the non-
interacting case where the perturbation is a one-particle operator and 
the self energy is (Keldysh) block-diagonal \cite{LL10} 
\begin{eqnarray}
\label{self}
\Sigma (t,t')=
\left( \begin{array}{cc}
V(t) & 0  \\ 
0    & V(t)  
\end{array}\right)\delta (t-t'). 
\end{eqnarray}
An impurity average effectively introduces interactions among the electrons 
and the self energy becomes different from Eq.~(\ref{self}). However, as long 
as no impurity average is performed, one merely has to deal with a one-
particle problem and the integral equation Eq.~(\ref{dyson}) together with 
Eq.~(\ref{self}) exactly determines the Green's function $G(t_{1},t_{2})$. 

Because of the linear relation $G^{T}+G^{\tilde{T}}=G^{>}+G^{<}$, a rotation 
to tridiagonal form can be performed in Eq.~(\ref{dyson}). One of the 
resulting equations will be used later, namely the one for the retarded 
Green's function 
$iG_{\alpha \beta }^{R}(t_{1},t_{2}):=
\theta(t_{1}-t_{2})\langle c_{\alpha }(t_{1})c_{\beta 
}^{\dagger}(t_{2})+c_{\beta}^{\dagger}(t_{2})c_{\alpha }(t_{1})\rangle$, 
which reads, using $\Sigma ^{R}(t,t')=V(t)\delta (t-t')$, 
\begin{eqnarray}
\label{dysonretarded}
G^{R}(t_{1},t_{2})=G^{0,R}(t_{1},t_{2})+\int dtG^{0,R}
(t_{1},t)V(t)G^{R}(t,t_{2}).
\end{eqnarray}

\subsection{Time-dependent field}
In all what follows we assume that the system is subject to a spatially 
homogeneous electric field which oscillates in time with a 
frequency $\Omega $ and is polarized in direction ${\bf e}$, \begin{equation} 
{\bf E}(t)\equiv{\bf e}E_{0}\cos(\Omega t). \label{photonfield} 
\end{equation} 
The associated vector potential  $ {\bf 
A}^{e}(t) = -({\bf e}E_{0}/\Omega )\sin(\Omega t)$
couples to the momentum ${\bf p}_{i}$ of the $i$-th
electron via
$
{\bf p}_{i}\to {\bf p}_{i}-{\bf A}({\bf x_{i}})-{\bf A}^{e}(t),
$
where ${\bf A}({\bf x_{i}})$ is included here to indicate that there can be other 
contributions to the vector potential, e.g from a static magnetic field
(we use units $\hbar=e=c=1$ throughout).
The {\em additional} energy through the electric field is
\begin{eqnarray}
\label{firstq}
H_{e}(t)= \frac{E_{0}}{\Omega }\sin(\Omega t)\sum_{i=1}^{N_{e}}{\bf e}{\bf 
v}_{i}+ N_{e}\Delta \Phi(t),\quad\Delta \Phi(t):=\frac{1}{2m^{*}\Omega 
^{2}}E_{0}^{2}\sin^{2}(\Omega t), \end{eqnarray} 
where $m^{*}$ is the bandmass, $m^{*}\bf{v}_{i}=  {\bf p}_{i}-{\bf A}({\bf 
x_{i}})$, and
$N_{e}$ the number of electrons in the 
system. We explicitely note that although the electric field ${\bf E}(t)$ is 
homogeneous, in an alternative gauge a corresponding {\em scalar potential} 
would be linear in the space coordinate, namely $\sim {\bf e x}\cos(\Omega t)$, 
an 'oscillating slanted surface'. 

By gauge invariance, the current density operator is 
$ {\bf j}({\bf 
x},t)={\bf j}_{0}({\bf x})-\frac{1}{m^{*}}{\bf A}^{e}(t)\rho({\bf x}),  
$ 
where ${\bf j}_{0}$ is the 
(paramagnetic) current density operator in the absence of the electric field, 
and $\rho({\bf x})$ the electronic density. 

In second quantization and in an basis of eigenstates $\{|\alpha \rangle\}$ 
with eigenenergies $\varepsilon_{\alpha }$, 
the electric field gives rise to an additional Hamiltonian 
\begin{eqnarray} 
\label{hamiltonian2}
H_{e}(t)&=& \frac{E_{0}}{\Omega 
}\sin(\Omega t)\sum_{\alpha \beta } \langle \alpha | v |\beta \rangle 
c_{\alpha }^{\dagger}c_{\beta }^{\phantom{\dagger}}+ 
\hat{N}\Delta \Phi(t) . 
\end{eqnarray} 
Here, $v=(1/m^{*}){\bf e}({\bf p-A(x)})$ is the component of the 
electron velocity operator in direction ${\bf e}$. Note that the coupling is 
effectively to the velocity and not to the current density operator since we 
assumed spatial homogeneity of the electric field. 

Care has to be taken when calculating Green's functions in  a
specific gauge, as has been pointed out by Bertoncini and 
Jauho \cite{BJ91}. Since only gauge-invariant quantities are meaningful for 
comparison with experiments, one has to formulate a  gauge-invariant theory.
In fact, a version $\tilde{G}$ of the Green's function which
is invariant under a gauge transformation of the electric field can be 
recovered from the Green's function $G$ in any specific gauge. In real space, 
this transformation is as follows \cite{BJ91}: if ${\bf R}=({\bf x+x'})/2$ and 
${\bf r} ={\bf x-x'}$ denote center-of mass and relative coordinates in real
space and $t=t_{1}-t_{2}$ and $T=(t_{1}+t_{2})/2$ the Wigner time variables,
\begin{equation}
\label{gauge}
\tilde{G}({\bf r},t,{\bf R},T)=\exp\left(iw({\bf r},t,{\bf R},T)\right)
G({\bf r},t,{\bf R},T),
\end{equation} 
where
\begin{equation}
\label{wdefinition}
w({\bf r},t,{\bf R},T):=\int_{-1/2}^{1/2}d\lambda 
\left [
\phi({\bf R}+\lambda {\bf r},T+\lambda t)-{\bf r}{\bf A}^{e}
({\bf R}+\lambda {\bf r},T+\lambda t)\right],
\end{equation}
and $\phi$ and ${\bf A}^{e}$ denote the static and the vector potential, 
respectively. Eq.~(\ref{gauge}) can be proofed directly by performing a gauge 
transformation ${\bf A}^{e}\to {\bf A}^{e}+\nabla \chi$, $\phi\to \phi-\partial_{t}
\chi$ and using the fact that the field operators in real space transform as
$\Psi\to \exp(i\chi) \Psi$ (note that we have set $e=\hbar=c=1$).
The factor $\exp(iw)$ becomes particularily simple in our case
of a spacially homogeneous electric field defined through $\phi = 0 $ and
${\bf A}^{e}(t)=-({\bf e}E_{0}/\Omega )\sin(\Omega t)$.
If we start in an arbitrary basis $\{|\alpha \rangle\}$ with wave functions
$\Psi_{\alpha }({\bf x})$, the gauge invariant object is
\begin{equation}
\label{gaugetrans}
\tilde{G}({\bf r},t,{\bf R},T)=\sum_{\alpha \beta }
\Psi_{\alpha}({\bf R+r/2}) \Psi_{\beta }^{*}({\bf R-r/2}) 
\exp\left(iw({\bf r},t,{\bf R},T)\right)G_{\alpha \beta }^{\Phi}(t,T),
\end{equation}
where $G_{\alpha \beta }^{\Phi}(t,T)$ is calculated in a specific gauge 
and basis ${|\alpha \rangle}$.

In the rest of this paper, we will always refer to a specific gauge and refer 
to Eq.~(\ref{gaugetrans}) for the final, gauge-invariant form. 
The importance of the phase factor $\exp(iw)$ depends on the 
physical situation and has been discussed, e.g., in \cite{BJ91,JJ96}. For 
convenience, we split off another phase factor in Eq.~(\ref{gaugetrans}).
The last term in the Hamiltonian Eq.~(\ref{hamiltonian2}) which is 
quadratic in $E_{0}$ couples to the total particle number 
operator $\hat{N}$ and thus gives rise to a phase factor
in the Green's function. The latter can be written as
\begin{eqnarray}
\label{phase}
G^{\Phi}(t_{1},t_{2})=e^{i\Phi(t_{1},t_{2})}G(t_{1},t_{2}),
\quad \Phi(t_{1},t_{2}):=N \int_{t_{1}}^{t_{2}} dt'
\Delta \Phi(t').                         
\end{eqnarray}  
where $N$ is the particle number and $G$ the Green's function for 
the same Hamiltonian Eq.~(\ref{hamiltonian2}) but without the last term 
which, as can been directly seen from Eq.~(\ref{hamiltonian2}),
shifts all energy levels $\varepsilon _{\alpha }$ by the same, but time 
dependent amount $\Delta \Phi(t).$ 

\section{Perturbation Theory in the impurity potential}
\label{ip}
In the preceding section we have derived the general equation 
Eq.~(\ref{dyson2}) which determines the Green's function of the system. 
As mentioned above, the splitting of the total Hamiltonian into perturbation 
part and unperturbed part defines the unperturbed Green's function $G^{0}$ 
which is at the base of any perturbation theory. In the following, we define 
the model Hamiltonian.
In this section, we start from 
an eigenstate basis of plane waves, i.e. $|\alpha \rangle=|{\bf k}\rangle$ 
where $\langle{\bf x}|{\bf k}\rangle\equiv \phi_{{\bf k}}({\bf x})=
(1/L^{d/2})\exp(-i{\bf kx})$ and $L^{d}$  
is the system volume ($L\to \infty$ in the thermodynamic limit).
We furthermore assume that there is no magnetic field, a
non--zero magnetic field can be included in 
the basis of exact scattering states treated in the next section. 
In the plane wave basis, the velocity matrix element is diagonal,
\begin{equation}
\label{velocitydia}
\langle \alpha |v|\beta \rangle=\delta _{\alpha ,\beta }\langle \alpha 
|v|\alpha \rangle\equiv \delta _{\alpha \beta }v_{\alpha },
\end{equation}
namely $v_{\alpha }=v_{{\bf k}}={\bf e}{\bf k}/m^{*}$. 
Although Eq.~(\ref{velocitydia}) is exact only for plane waves, we keep the 
notation as general as possible and use greek letters labeling the 
eigenstates. The following results are then also valid for a general 
eigenstate basis with the approximation that non-diagonal elements of the 
velocity $\langle \alpha |v|\beta \rangle$ are zero.    
Because of Eq.~(\ref{velocitydia}), the Hamiltonian (system + electric field)
is
\begin{equation}
\label{hamepstime}
H_{0}(t)=\sum_{\alpha }\varepsilon _{\alpha }(t)c^{\dagger}_{\alpha }c
^{\phantom{\dagger}}_{\alpha };
\quad \varepsilon _{\alpha }(t)\equiv \varepsilon _{\alpha }+\frac{E_{0}}
{\Omega }v_{\alpha }\sin(\Omega t),        
\end{equation}
where $\varepsilon _{\alpha }$ is the energy of state $\alpha $.

The static potential $V({\bf x})$ now gives rise to an 
additional part to the total Hamiltonian which reads 
\begin{equation}
\label{fullh}
H(t)=H_{0}(t)+V,\quad V=\sum_{\alpha \beta }V_{\alpha \beta }
c_{\alpha}^{\dagger}c_{\beta }^{\phantom{\dagger}}.
\end{equation}
At this stage, $V$ is not yet specified further. Depending on the physical 
situation, it describes a single impurity, a distribution of random 
scatterers, a double barrier etc.  Note that 
according to the above, in Eq.~(\ref{fullh}) $\alpha $ denotes a plane wave 
while in the corresponding 'diagonal approximation' case for arbitrary basis 
$\alpha $ the potential term $V$ does not appear and simply $H(t)=H_{0}(t)$.

\subsection{Dyson equation in arbitrary strong electric fields}
We define an interaction picture with respect to the unperturbed 
part $H_{0}$ of the Hamiltonian according to
\begin{equation}
\label{ipicture}
\hat{c}_{\alpha }(t)=U^{\dagger}_{0}(t,t_{0})c_{\alpha }U_{0}(t,t_{0}),
\quad U_{0}(t,t_{0})=T\exp\left( -i\int_{t_{0}}^{t}dt' H_{0}(t')\right).
\end{equation}
The unperturbed time-ordered Green's function then is 
\begin{eqnarray}
\label{unperturbedg}
iG^{T,0}_{\alpha \beta }(t_{1},t_{2})&=&
\langle T\hat{c}_{\alpha }(t_{1})\hat{c}_{\beta 
}^{\dagger}(t_{2})\rangle=\nonumber\\ 
&=&e^{-i\int_{t_{2}}^{t_{1}} dt' \varepsilon _{\alpha }(t')}\left[ \theta
(t_{1}-t_{2})(1-f_{\alpha })-\theta(t_{2}-t_{1})f_{\alpha }\right]\delta 
_{\alpha \beta }, \qquad f_{\alpha }=
\langle c_{\alpha}^{\dagger}c_{\alpha }^{\phantom{\dagger}}\rangle
\end{eqnarray}
and correspondingly for the other Green's functions.
The average $\langle\rangle$ gives rise to  a distribution 
function $f_{\alpha }$ which not necessarily coincides with the Fermi 
distribution. Note that the value of the fixed time $t_{0}$ in 
Eq.~(\ref{ipicture}) is not relevant for $G$. The Green's function
Eq.~(\ref{unperturbedg})
can be related to the free ($V=0$) Green's function without electric field
($E_{0}=0$), which in the following is called $G^{free}$. Namely, one has
for the time-ordered part
\begin{equation}
\label{gfree}
G^{T,0}_{\alpha \beta }(t_{1},t_{2})=e^{ig_{\alpha }[\cos(\Omega t_{1})
-\cos(\Omega t_{2})]} G^{T,free}_{\alpha}(t_{1}-t_{2})\delta _{\alpha \beta },
\qquad g_{\alpha }:=\frac{E_{0}v_{\alpha }}{\Omega ^{2}}
\end{equation}
and correspondingly for the other Keldysh-block components.
The effect of the electric field thus is incorporated in the 
phase factor in front of $G^{T,free}$ in Eq.~(\ref{gfree}). Note that this 
phase factor, in contrast to the 'global' phase $\Phi(t_{1},t_{2})$, 
Eq.~(\ref{phase}), depends 
on the quantum number $\alpha $ which will be important when
scattering, i.e. transitions between different states $\alpha \to \beta $, is 
considered. 
            
To get into contact with the Fourier representation above, we use a 
decomposition into Bessel functions according to 
$\exp[iz\cos (\Omega t)]=\sum_{n}i^{n}J_{n}(z)\exp(in\Omega t)$ (the 
sum is over all integers $n$). We obtain                        
\begin{equation}
\label{gbessel}
G^{0}_{\alpha \beta }(\omega ,N)=i^{N}\sum_{n}J_{n}(g_{\alpha })J_{N-n}(-g_{\alpha })
G^{free}_{\alpha }(\omega +[n-N/2]\Omega )\delta _{\alpha \beta }
\end{equation}
for the Keldysh Green's function matrix. Note that $G^{free}$ depends on one 
Fourier variable only since the corresponding $G^{free}(t_{1}-t_{2})$ is a 
function of the time difference only. We note again that the eigenstate basis 
in this section consists of plane waves $|\alpha\rangle =|{\bf k}\rangle$ 
with momentum ${\bf k}$. Only in the 'diagonal approximation' model, where 
one neglects the potential term $V$ in Eq.~(\ref{fullh}), the results of this 
section can be transfered to an arbitrary eigenstate basis. Furthermore, in 
this section $G^{0}$ denotes the Green's function belonging to the {\em time-
dependent} Hamiltonian $H_{0}(t)$, Eq.~(\ref{hamepstime}). It contains the 
electric field exactly to all orders of the coupling parameter $g_{\alpha 
}$, Eq.~(\ref{gfree}), which appears in the argument of the Bessel functions. 

We now proceed to set up a systematic approximation scheme that allows to 
calculate the Fourier components $G_{\alpha \beta }(\omega ,N)$ of the 
Green's function belonging to the time-dependent Hamiltonian $H(t)$, 
Eq.~(\ref{fullh}). In particular, the effect of the time dependent field 
will be fully incorporated through the unperturbed Green's 
functions $G^{0}$, Eq.~(\ref{gbessel}). 

Since the perturbation potential $V$ is time-independent, the selfenergy
in Eq.~(\ref{dyson2}) is 
\begin{equation}
\Sigma (\omega ,N)=S \cdot\delta _{N,0},\quad 
S=
\left( \begin{array}{cc}
V & 0  \\ 
0    & V  
\end{array}\right).
\end{equation}  
Thus, 
\begin{eqnarray}
\label{dyson3}
G(\omega ,N)&=&G^{0}(\omega ,N)+\sum_{N_{1}}G^{0}\left(\omega 
+\frac{N_{1}-N}{2}\Omega ,N_{1}\right)S 
G\left(\omega 
+\frac{N_{1}}{2}\Omega ,N-N_{1}\right).
\end{eqnarray}
One immediately sees the fundamental difficulty in Eq.~(\ref{dyson3}): Even for 
obtaining only the Fourier component $G(\omega ,N=0)$, corresponding to an 
average of the 'center-of-mass' time coordinate, the $N_{1}$-sum couples the 
different $N$--components in Eq.~(\ref{dyson3}).

An obvious approximation would be to neglect all $N\ne 0$ components at all, 
i.e. to neglect the center-of-mass time coordinate. Indeed, approximations of 
this kind have been used recently (we call it 'fast' approximation in the 
following), namely in the so-called 'non-adiabatic regime' of resonant 
tunneling \cite{AIP96} where the frequency $\Omega $ of the time dependent 
field is much larger than the inverse tunneling time. Another example is the 
neglect of higher order Fourier components in the calculation of Green's 
functions for Wannier Stark systems \cite{HH96}. There, the approximations 
were argued to work well for frequencies $\Omega $ larger than the bandwidth 
of the systems (the letter is proportional to the hopping matrix element of 
the corresponding tight-binding Hamiltonian). 
The extreme advantage of those approximations is that in a relatively simple 
manner the effect of the time dependent field can be incorporated by 
essentially calculating expressions where the free Green's function 
$G^{free}(\omega )$ is replaced by $G^{0}(\omega,N=0)$, Eq.~(\ref{gbessel}). 

In the general situation of an (arbitrary) disorder potential or an arbitrary 
eigenstate basis, it is not clear if an analogy to the fast approximation 
exists. Furthermore, the validity of the latter has not been investigated 
systematically to our knowledge. In the following, we develop a scheme for 
such an investigation. 
The formalism set up below is completely general and can be applied to a large 
range of other situations, too. This opens the possibility to go beyond the 
'fast approximation' in a general scattering situation.

\subsection{Fast approximation and its first corrections}
\label{fastapproximation}
We start calculating the $N=0, \pm 1$ components of
$G(\omega,N)$, Eq.~(\ref{dyson3}), i.e. the 'fast' approximation  which 
neglects the 'center--of--mass' time coordinate $T$, and its first 
corrections. 
Introducing the  convenient notation 
\begin{equation}
\label{convenient}
g_{l,n}^{(0)}:=G^{(0)}\left(\omega +\frac{l}{2}\Omega ,n\right),
\end{equation}
we obtain from the terms $N_{1}=0,\pm 1$ in Eq.~(\ref{dyson3})
the approximate solution
\begin{equation}
g_{l,n}\approx g_{ln}^{0}+g^{0}_{l-n,0} S  g_{l,n}+g^{0}_{1+l-n,1} S  g_{1+l,n-1}+
g^{0}_{-1+l-n,-1} S  g_{l-1,n+1}.
\end{equation}
Here, all quantities are (Keldysh block-) matrices. 
The problem of 
determining the (time-average) component $g_{0,0}=G(\omega ,N=0)$ is pure 
algebra and one obtains 
\begin{eqnarray}
\label{g00}
G(\omega ,N=0)&\approx&(1-\tilde{g}^{0}S )^{-1}\tilde{g}^{0}\nonumber\\
\tilde{g}^{0} &=&g^{0}_{0,0}+g^{0}_{1,1}S (1-g_{2,0}^{0}S )^{-
1}g^{0}_{1,-1}+ g^{0}_{-1,-1}S (1-g_{-2,0}^{0}S )^{-1}g^{0}_{-1,1}. 
\end{eqnarray} 
Equivalently, other components than $N=0$ might be calculated.
Note that Eq.~(\ref{g00}) has the standard form of the formal solution of 
the time-independent Dyson equation for a self energy $S $ (which for our 
case of a static potential $V$ is given by $V$ itself), 
\begin{equation} 
\label{dysonnolight} 
G(\omega )=(1-G^{free}(\omega )S )^{-1}G^{free}(\omega ). 
\end{equation} 
However, the 'unperturbed' Green's function
$g_{0,0}^{0}$ is substituted by the more complicated expression 
$\tilde{g}^{0}$. In fact, having $g_{0,0}^{0}$ instead of  $\tilde{g}^{0}$
in Eq.~(\ref{g00}) is exactly the 'fast approximation', namely
\begin{equation}
\label{g00fast}
G(\omega ,N=0)\approx(1-G^{0}(\omega,N=0)S )^{-1}G^{0}(\omega ,N=0).
\end{equation}
Eq.~(\ref{g00fast}) is the zeroth order approximation for 
Eq.~(\ref{dyson3}) with respect to the Fourier index $N$. In particular, the 
dependence on the 'center-of-mass' time coordinate $T$, Eq.~(\ref{fourier}) 
is completely neglected here.

\subsection{Exact solution}
The perturbative calculation of the $N=0,\pm 1$--components of
$G(\omega ,N)$ above
suggests that there is a general solution of the Dyson equation 
type which holds to all orders. Eq.~(\ref{dyson3}) can be solved 
formally to {\em all orders} in the self energy and for all components $N$. 
The procedure for this is very similar to the derivation of  
the Floquet-matrix in the exact eigenstate basis below. Here, we derive the 
exact result in the plane wave basis. It is the generalization of the Dyson 
equation solution for a time--independent scattering problem to the case of 
an additional time-dependent electric field and will be the 
starting point for the subsequent numerical truncation scheme. 

We first rewrite 
Eq.~(\ref{dyson3}) as 
\begin{equation}
g_{ln} = g^{0}_{ln}+\sum_{n'}g^{0}_{l+n',n'+n}Sg_{l+n+n',-n'},
\end{equation}
where we shifted the $n'$-summation index. We introduce 
a matrix $\gamma $ with elements $\gamma 
_{rs}=g_{ln}\equiv g_{r+s,s-r}$, together with a matrix
$\gamma ^{0}$ with the corresponding elements of $g^{0}$. We perform an
index transformation
\begin{equation}
\label{indextrafo}
r=(l-n)/2,\quad s=(l+n)/2,
\end{equation}
which yields 
\begin{equation}
\gamma _{rs}=\gamma ^{0}_{rs}+\sum_{n'}\gamma ^{0}_{rn'}S\gamma _{n's}.
\end{equation}
Therefore, in the space of the $(r,s)$-indices, the solution of the matrix 
equation becomes
\begin{equation}
\gamma =(1-\gamma ^{0}S)^{-1}\gamma ^{0}.
\end{equation}
Explicitely, the matrix $\gamma $ has the form
\begin{eqnarray}
\label{exactpot}
\gamma =\left( \begin{array}{ccccc}
... & & & &   \\ 
    &1-g_{-20}^{0}S &-g^{0}_{-11}S  &-g_{02}^{0}S &...   \\
... &-g^{0}_{-1-1}S  &1-g_{00}^{0}S &-g^{0}_{11}S   &...   \\
    &-g^{0}_{0-2}S  &-g^{0}_{1-1}S &1-g_{20}^{0}S  &...  \\
... & & & &.... \\
\end{array}\right)^{-1}\times \gamma^{0}. 
\end{eqnarray}
In particular, the matrix $(1-\gamma^{0}S)$ is fully occupied. 

Since $\gamma $ is just another representation of the original matrix 
$g$, $\gamma_{rs}=g_{r+s,s-r}$, Eq.~(\ref{exactpot}) solves the Dyson 
equation Eq.~(\ref{dyson3}) exactly.
Note that the elements of the matrix $\gamma $ are (Keldysh) matrices
themselves. This tensor structure is 
analogous to the formal solution in the exact eigenstate basis below, here
we started from a plane wave  eigenstate basis
$|{{\bf k}\rangle}$. While $\gamma ^{0}$ is diagonal in this basis, the self
energy $S$ in general is not diagonal and depends on $k$ and $k'$ for a given 
static potential.

In the usual time-independent case, solving the Dyson equation 
Eq.~(\ref{dysonnolight}) for an 
arbitrary potential requires the inversion of a matrix of dimension
$D$, where $D$ is the dimension of the eigenstate basis (which is finite 
in any numerical calculation).
In the present time-dependent case, the effective dimension of the matrix 
$\gamma$ is $D$ times $n_{ph}$, where $n_{ph}$ is the number of  
'photoblocks' in $(1-\gamma ^{0}S)$, e.g. $n_{ph}=3$ for the three blocks 
per row in Eq.~(\ref{exactpot}). Since 
there are infinitely many Fourier components of the Green's function, in the 
exact solution Eq.~(\ref{exactpot}), $n_{ph}=\infty$, and $\gamma $ is of 
infinite dimension even if the dimension $D$ of the original Hilbert space is 
finite. It turns out, however, that it is sufficient to consider only a relatively 
small number of these blocks in practical calculations. In principle, the 
numerical effort increases rapidly with the number $n_{ph}$. Fortunately, 
the examples below show that convergence is reached quickly by truncating 
$\gamma $ at relatively small $n_{ph}$. We also stress that this 'truncation 
method' is to infinite order in the static potential $V$ {\em and} the 
electric field $E_{0}$. It is perturbative in the higher {\em Fourier 
components} (large $n$) for the center--of--mass time coordinate $T$ in 
$g^{0}_{ln}$ in the off--diagonal in Eq.~(\ref{exactpot}). 
                        
Since Eq.~(\ref{exactpot})  is the exact solution of Eq.~(\ref{dyson3}), one 
easily recovers the above approximations: The 'fast approximation' 
Eq.~(\ref{g00fast}) is obtained by cutting the matrix $(1-\gamma^{0}S)$ such 
that only the $(r=0,s=0)$, i.e. the 'central' element is retained. In $(r,s)$ 
space the (infinite) matrix becomes  a number and the inversion is trivial. 
The next approximation Eq.~(\ref{g00}) that retains only the $N=0,\pm 1$ 
components can be verified similarly by inversion of the corresponding $3 
\times 3$ matrix in $(r,s)$-space. 

Eq.~(\ref{exactpot}) offers the possibility of a systematic investigation of 
the 'truncation method', in particular of approximations like Eq.~(\ref{g00}) 
or Eq.~(\ref{g00fast}).                         

\subsection{Numerical examples}
In the following, we apply the formalism to calculate 
the {\em time-averaged} Green's function $G(\omega ,N=0)$ numerically, using 
Eq.~(\ref{exactpot}) in two different potentials $V$. 
\subsubsection{Constant potential}
In the first model, we use a constant self energy $V=const$. 
Physically, this is a trivial, nevertheless very instructive
situation since it offers comparison to an exact solution.
That is, the constant shift of the energy scale by $V$ corresponds to the 
exact result 
\begin{equation}
\label{exact}
G_{k}(\omega ,N)=G_{k}^{0}(\omega -V,N)
\end{equation}
which has to be reproduced by inverting the full Dyson equation, 
Eq.~(\ref{exactpot}). Here, $G_{k}^{0}$ is the unperturbed $(V=0)$ Green's 
function defined according to Eq.~(\ref{gbessel}).

The result of the numerical inversion is shown for the $N=0$--component of 
the corresponding spectral function component which we defined from the 
Green's function as   
\begin{equation} 
\label{spectralfunction}
A_{k}(\omega,N=0):=-\frac{1}{\pi}{\rm 
Im}\,G^{R}_{k}(\omega ,N=0), 
\end{equation} 
through the retarded Green's function. The numerical calculation here is regarded as a check for the 
consistency and convergence of the formalism as such. Therefore, for reasons of 
simplicity we did not calculate the full 'physical' spectral function 
$A_{k}(\omega )$ \cite{JJ96} which involves the additional phase factor $w$, 
Eq.~(\ref{gaugetrans}). 
Note that because of Eq.~(\ref{dysonretarded}), we only have to 
consider retarded functions in the Dyson equation. In particular, in our 
plane wave basis the elements $1-G^{0}_{l,n}S=1-G^{0}_{l,n}V$ are (complex) 
numbers and not matrices since we are working in Fourier space and not in 
real space here. 

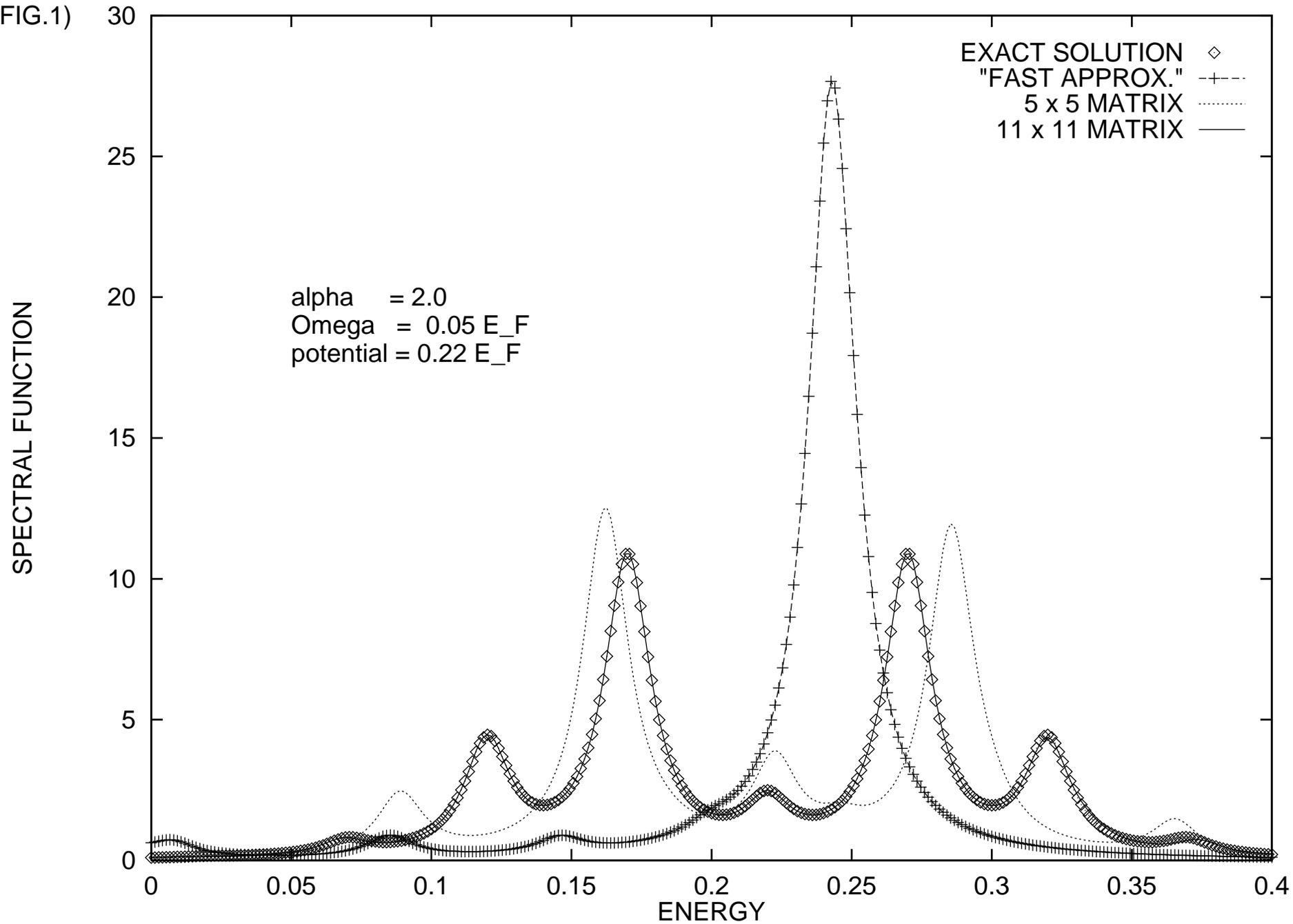
\begin{figure}[]
\unitlength1cm
\caption[]{\label{greenb}The $N=0$ spectral function component
Eq.~(\ref{spectralfunction}) in units of the inverse Fermi energy 
$E_F^{-1}$ at the Fermi wave vector $k=k_F$. The model system 
describes a constant potential $V$ and a time-dependent electric field of 
frequency $\Omega $. The dimensionless parameter $\alpha $ 
regulates the coupling to the electric field, energies are in units of $E_F$. 
The exact solution is shown together with the result from the
'fast approximation', Eq.~(\ref{g00fast}), and the result from the numerical inversion of 
Eq.~(\ref{exactpot}). Already for $n_{ph} =11$, practically for 
all energies coincidence is reached with the exact result Eq.~(\ref{exact}).} 
\end{figure}

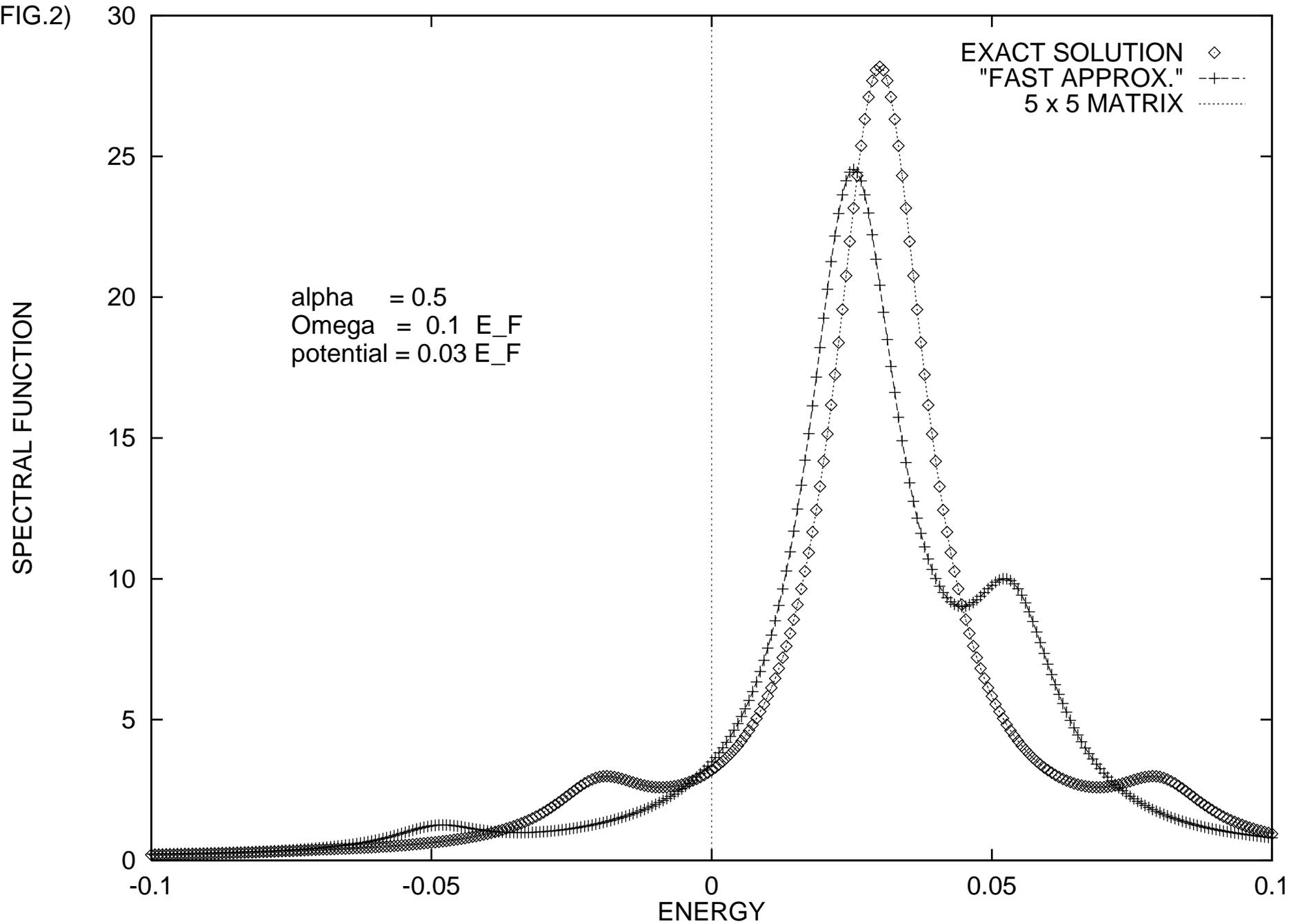
\begin{figure}[]
\unitlength1cm
\caption[]{\label{greena}Same as in Fig.~(\ref{greenb}) but for weaker 
coupling $\alpha $ and smaller potential $V$. Coincidence is reached for 
an even smaller matrix Eq.~(\ref{exactpot}) in this case.} 
\end{figure}

The results of the numerical inversion of Eq.~(\ref{exactpot}) are shown in 
Fig.~(\ref{greenb}) and Fig.~(\ref{greena}). We  fixed the wavevector 
$k=k_{F}$,  energies are measured in units of the Fermi energy
$E_{F} =(\hbar k_{F})^{2}/2m^{*}$. The dimensionless parameter $\alpha $ 
regulating the coupling to the electric field of strength $E_{0}$ is 
defined as 
\begin{equation}
\alpha=E_{0}k_{F}/(m^{*}\Omega ^{2}).
\end{equation}
For a potential 
$V=0.22 E_{F}$ and strong coupling, $\alpha =2$, the central peak of the 
spectral function is strongly suppressed: its main weight comes from the 
Bessel functions $J_{0}^{2}$, on the other hand, the product $\alpha k$ is 
close to the first zero of the latter. Side-peaks appear at values of the 
energy $\omega $ shifted by multiples $n\Omega $. The exact result 
Eq.~(\ref{exact}) is excellently reproduced by inversion of the $11\times 11$ 
'photo'block matrix Eq.~(\ref{exactpot}), while the 'fast approximation' 
Eq.~(\ref{g00fast}) gives a totally wrong result. In contrast to this, 
already the inversion of the $5 \times 5$ matrix qualitatively reproduces the 
exact result. The situation is less dramatic for smaller coupling $\alpha 
=0.5$ and smaller potential $V=0.03 E_{F}$. There, the 'fast approximation'
Eq.~(\ref{g00fast}) is qualitatively roughly correct whereas the exact result 
is already obtained from the $5\times 5$ matrix inverse.                             

\subsubsection{Resonant tunneling}
In the second example, we investigated a physically non-trivial situation, 
namely the transmission of an electron through a double-barrier structure.
The model is a one-dimensional lattice with $j_{max}$ sites and a potential
\begin{eqnarray}
S=V=|x_{1}\rangle V_{1}\langle x_{1}|+|x_{2}\rangle V_{2}\langle x_{2}|.
\end{eqnarray}
For $V_{1},V_{2}>0$, this potential has a resonant structure in its transmission coefficient 
and can be regarded as a simple version of a model simulating a quantum well 
double barrier. The Dyson equation Eq.~(\ref{dyson3})
has to be transformed to real space to obtain the matrix elements
$\langle x | G(\omega ,N)|x'\rangle$ of the full Green's function in real 
space.
Changing from the original plane-wave basis to real space in particular
involves a Fourier transformation in the Green's function $G^{0}$, 
\begin{equation}
\label{fourierfree}
G^{0}(x-x';\omega ,N)=\frac{1}{2\pi}\int dk e^{ik(x-x')}G_{k}^{0}(\omega ,N).
\end{equation}
In fact, the integral Eq.~(\ref{fourierfree}) is tedious and we found an 
analytical form for the retarded Green's function 
only for small couplings $x_{0}=E_{0}/m^{*}\Omega ^{2}$, namely 
\begin{eqnarray}
G^{0,r}(x-x';\omega ,N)&=&
\left[ {\rm sgn}(x-x')i \right]^{N}\sum_{n}J_{n}(x_{0}\sqrt{z})
J_{N-n}(-x_{0}\sqrt{z})\frac{-i}{2\sqrt{z}}e^{i\sqrt{z}|x-x'|},\nonumber\\
z&:=& \omega +(n-N/2)\Omega ,\quad 2x_{0}<|x-x'|
\end{eqnarray}
For smaller $|x-x'|$, we resorted to a numerical evaluation of the integral.
We defined the coefficient
\begin{equation}
\label{transmission}
T(\omega ):=\left| \frac{G^{r}(x,x',\omega ,N=0)}{G^{0,r}(x-x',\omega ,N=0)}
\right|^{2}
\end{equation}
from the time-averaged $N=0$--components of the Green's functions which is 
independent of $x,x'$ for $x<x_{1}$ and $x'>x_{2}$, i.e. transmission of an 
incoming electron (from left) through both potentials at sites $x_{1}$ and 
$x_{2}$. Strictly speaking, since $T(\omega )$ is defined by the product of 
the time-averaged $G$'s and not the time-average of the product $G G^{*}$, 
$T(\omega )$ deviates from the (physical) time-averaged transmission 
coefficient. Here as above we restrict ourselves, however, to the simpler 
quantity Eq.~(\ref{transmission}) to clarify the convergence of the numerical 
method to determine $G$. We chose $j_{max}=12$ and a distance $x_{2}-
x_{1}=2a$, where $a$ is the lattice constant serving as length scale and  
$E_{a}=\hbar^{2}/ 2m^{*}a^2$ for the energy scale. The number $n_{ph}$ of 
'photo-blocks' in the matrix $(1-
\gamma^{0}S)$ again determines the total size of the matrix to be inverted. 
Each block itself is a $j_{max} \times j_{max}$--matrix, the total dimension 
of the matrix therefore being $j_{max} \times  n_{ph}$. 

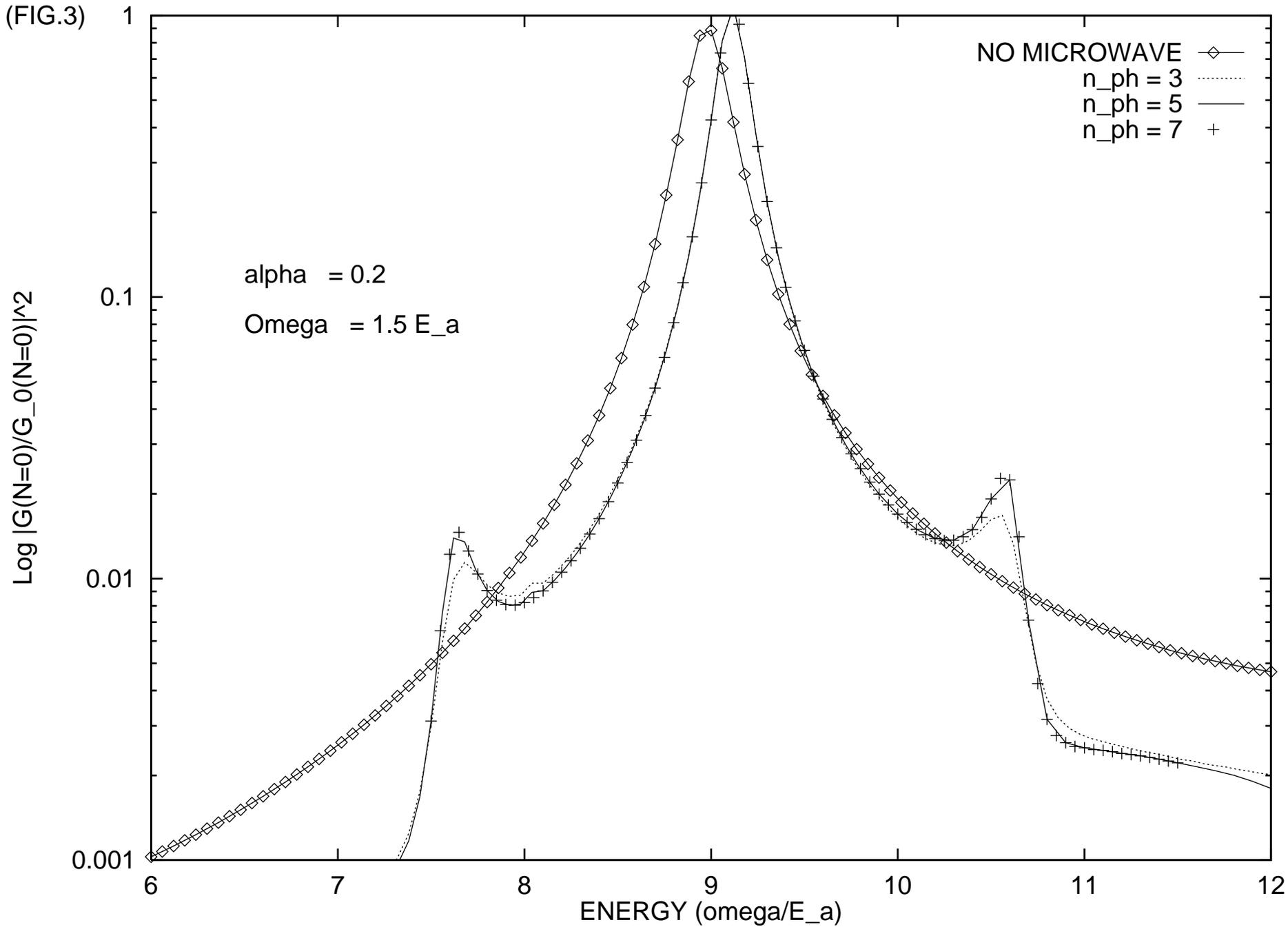
\begin{figure}[] 
\unitlength1cm 
\caption[]{\label{greent}Transmission coefficient Eq.~(\ref{transmission}) 
through a one-dimensional double barrier model in a time-dependent 
electric field with frequency $\Omega $ and coupling parameter $\alpha $. 
Again, the Green's function has been calculated numerically for different 
numbers $n_{ph}$ of 'photoblocks' in order to check the convergence of the 
results from inverting the truncated matrices Eq.~(\ref{exactpot}). The 
system is defined by a 1d tight-binding model with 12 sites. The distance of 
the two potentials of strength $V=20 E_{a}$ defining the barrier is $2a$ with 
$a$ the lattice constant; energies are in units of $E_{a}\equiv\hbar^{2}/ 
2m^{*}a^2$. Site peaks in the transmission coefficient occur a energies 
shifted by $\pm \hbar\Omega $ from the central transmission peak.} 
\end{figure}

Fig.~(\ref{greent}) shows the coefficient $T(\omega )$ for three different 
sizes of the matrix Eq.~(\ref{exactpot}). One reckognizes that already for 
$n_{ph}=5$ convergence is reached practically for all values of the energy $\omega $. The 
side-peaks to the left and to the right of the transmission maximum 
correspond to the first 'side-bands' $\pm \hbar \Omega$ for tunneling through 
the double barrier and have been found previously by I\~{n}arrea et al. 
\cite{IPT94}. 

We conclude that the numerical inversion of Eq.~(\ref{exactpot})
shows the extremely good convergence of the results from truncating 
Eq.~(\ref{exactpot}), in contrast to the approximation Eq.~(\ref{g00fast}). 
This in particular indicates that one has to be very careful when using 
approximations where Fourier components of the Green's function are
neglected. We rather believe that our 
calculation shows that in general one has to consider at least a certain
number of  components $G(\omega,N)$, $N\ne 0$, from the very beginning. In 
the original representation of $G(t_{1},t_{2})$, Eq.~(\ref{fourier}), this 
means that not only the relative time coordinate $t_{1}-t_{2}$ but also the 
'center-of-mass' time coordinate $T$ is important. 

\section{Time dependent field as perturbation}
\label{fieldperturbation}
The basic advantage of the above approach is its exactness to 
all orders in the electric field which {\em a priori} is incorporated in the 
free Green's function $G^{0}$. The exact result Eq.~(\ref{exactpot}) implies 
that one solves the {\em static} scattering problem and the effect of the 
electric field simultaneously.        By inverting the truncated matrix  one 
sums up the static potential $V$ to infinite order. Still,  
higher Fourier component terms $Sg_{rs}^{0}$ proportional to $V$ in the off-
diagonal of Eq.~(\ref{exactpot}) are neglected. On the other hand, there are situations 
where one wishes to make no approximations concerning the static 
potential, in particular in systems where localization plays a role. 
Furthermore, in cases where the static potential is complicated, practical 
calculations become tedious in the plane wave basis. In fact, if the static 
scattering problem is already solved (say by numerical diagonalization), the 
basis of the static potential eigenstates provides the ideal starting point. 

In the following, we therefore take a different point of view and consider 
the electric field as a perturbation while starting from the exact basis of 
eigenstates of the system, i.e. the scattering states of 
the static potential $V$.
In presence of an additional, static magnetic field ${\bf B}$, these are the 
scattering states of the static potential in presence  of ${\bf B}$.
As above, a formally exact expression for the Green's function is derived 
which is closely related to the Floquet state formalism \cite{Shi65}. As an 
application, we derive an exact expression for the dephasing time due to the 
time-dependent electric field expressed in terms of the AC-conductivity of 
the system in absence of the electric field. This expression in particular is 
valid near or at a metal-insulator transition. 

We start from a Hamiltonian
\begin{eqnarray}
\label{hamiltonian20}
H(t)&=&H_{0}+M(t)=\sum_{\alpha} \varepsilon _{\alpha }
c_{\alpha}^{\dagger}c_{\alpha}^{\phantom{\dagger}}+
\sum_{\alpha \beta }M_{\alpha \beta }(t)
c_{\alpha}^{\dagger}c_{\beta }^{\phantom{\dagger}}\nonumber\\
M_{\alpha \beta }(t)&=&M_{\alpha \beta }\sin(\Omega t)= \frac{E_{0}}{\Omega 
}\langle \alpha | v |\beta \rangle \sin(\Omega t). 
\end{eqnarray} 
The unperturbed part $H_{0}$ is now time-independent and describes the 
non-interacting system in the exact eigenstate basis, labeled with the index
$\alpha $. An additional advantage is that the unperturbed quantities are now 
defined in an equilibrium system in contrast to the case in the preceding 
section. There, $H_{0}(t)$ was time dependent and thus even the unperturbed 
system not in equilibrium. 

The  matrix element $M_{\alpha \beta }$ describes transitions from 
states $\alpha $ to states $\beta $ and couples to the velocity
matrix element $\langle \alpha |v|\beta \rangle$ in polarization direction of 
the electric field.
In Fourier representation, the free Green's function $G^{0}$ belonging to 
$H_{0}$ is diagonal and depends on the time difference only, while the self 
energy $\Sigma (t,t')=\delta (t-t')\sin(\Omega t)\Sigma $ has components
$N=\pm 1$,
\begin{eqnarray}
G^{0}_{\alpha \beta }(\omega ,N)&=&\delta _{N,0}\delta 
_{\alpha \beta }G^{0}_{\alpha }(\omega )\nonumber\\
\Sigma _{\alpha \beta }(\omega ,N)&=&
\Sigma _{\alpha \beta }
[\delta _{N,1}-\delta_{N,-1}],\quad
\Sigma _{\alpha \beta }\equiv
\frac{1}{2i}W_{\alpha \beta }
\equiv\frac{1}{2i}
\left( \begin{array}{cc}
M_{\alpha \beta } & 0  \\ 
0    & M_{\alpha \beta }\end{array}\right). 
\end{eqnarray}
Using the general Eq.~(\ref{dyson2}) one obtains 
\begin{eqnarray}
\label{dysonexact}
G(\omega ,N)=G^{0}(\omega )\delta _{N,0}+G^{0}\left(\omega -\frac{N\Omega }{2}\right)
\Sigma \left[G\left(\omega +\frac{\Omega }{2},N-1\right)-
G\left(\omega -\frac{\Omega }{2},N+1\right)\right].
\end{eqnarray}
For convenience,  we again use the notation Eq.~(\ref{convenient})
$g_{l,n}=G(\omega +l\Omega /2,n)$, furthermore
$g_{l}^{0}=G^{0}(\omega +l\Omega /2)$,
so that Eq.~(\ref{dysonexact}) becomes
\begin{equation}
\label{recursion}
g_{l,n}=\delta _{n,0}g^{0}_{l}+g^{0}_{l-n}\Sigma [g_{l+1,n-1}-g_{l-1,n+1}].
\end{equation}
We note that again all quantities are Keldysh block matrices, each block 
being a matrix in the indices $\alpha ,\beta $ that label the {\em exact} 
eigenstates of $H_0$ (static scattering potential plus magnetic field). 
The free Green's function $G^{0}$ in this chapter 
thus refers to the time-independent part $H_{0}$ only and is diagonal in the 
eigenstate basis.

\subsection{Exact formal solution}
\label{exactformal}
We use the index transformation Eq.~(\ref{indextrafo}) $r =(l-n)/2, s 
=(l+n)/2, r ,s =0,\pm 1, \pm 2,...$ which allows for a formal solution of 
Eq.~(\ref{recursion}) and shows the relation to the Floquet state theory. 
Again, we define the matrices
\begin{eqnarray}
\label{gamma}
\Gamma _{r s }  &:=&g_{r +s ,s -r }\equiv 
g_{l,n}\nonumber\\
\Gamma _{2r }^{0}&:=&g^{0}_{l-n},\quad
\Gamma _{r s }^{0}:=\delta _{r s }\Gamma _{2r }^{0}=
\delta _{n,0}g_{l}^{0}.
\end{eqnarray}
Note that in contrast to the previous section, $\Gamma $ and $g$ now refer to 
the exact eigenstate basis. The requirement that both $r $ and $s $ are 
integer numbers in fact restricts the value of $l$: Even $n=s-r$ requires 
even $l=s+r$ and odd $n$ requires odd $l$. From the definition of 
$g_{ln}=G(\omega +l/2\Omega ,n)$ one reckognizes, however, that the index $l$ 
is redundant in the sense that {\em one} value of $l$ is sufficient to obtain 
the full information of the function $G(\omega ,n)$. This kind of redundance 
is well-known from the Floquet state formalism \cite{Shi65} to which the 
analogy becomes evident in the following. 

From Eq.~(\ref{recursion}), we obtain a corresponding equation in the
$r ,s -$ space, using 
$g_{l\pm 1,n\mp 1}=\Gamma _{r \pm 1,s }$, 
\begin{eqnarray}
\Gamma _{r s }=\Gamma _{rs }^{0}+\Gamma ^{0}_{2r}
\Sigma \left[ \Gamma _{r +1,s }-\Gamma _{r -1,s 
}\right].
\end{eqnarray}
After introducing
\begin{eqnarray}                   
\sigma _{r r '}=(\delta _{r +1,r '}-
\delta _{r -1,r '})\Sigma ,
\end{eqnarray}
this can be written in the matrix form 
\begin{eqnarray}
\label{formalsolution}
\Gamma =\Gamma ^{0}+\Gamma ^{0}\sigma \Gamma =\left[(\Gamma ^{0})^{-1}-\sigma 
\right]^{-1}.
\end{eqnarray}
Eq.~(\ref{formalsolution}) is the formal solution of the problem to determine 
the Green's function for the Hamiltonian Eq.~(\ref{hamiltonian20}). The 
structure of the matrix $\Gamma $ is as follows:
Since $\Gamma^{0}$ is diagonal in $r$ and  $s$, one has $(\Gamma^{0})^{-
1}_{r s }=\delta _{r s }(g^{0}_{l=2r ,n=0})^{-1}=
\delta _{r s }G^{0}(\omega +r \Omega )^{-1}$.
Therefore, the matrix $\Gamma $ is the inverse of a tridiagonal matrix,
\begin{eqnarray}
\label{floquet}
\Gamma =\left( \begin{array}{cccccc}
... & & & & &  \\ 
    &G^{0}(\omega -\Omega )^{-1} &-\Sigma  & &\mbox{\huge 0} &  \\
    &\Sigma  & G^{0}(\omega)^{-1}&-\Sigma  & &  \\
    & &\Sigma  &G^{0}(\omega +\Omega )^{-1} &-\Sigma  &  \\
    &\mbox{\huge 0} & &\Sigma  &G^{0}(\omega +2\Omega )^{-1} & -\Sigma  \\
    & & & &\Sigma  &...  
\end{array}\right)^{-1}, 
\end{eqnarray}
i.e. an infinite tridiagonal block matrix in the space of 'photon numbers'. Note that the 
elements of $\gamma$ are (Keldysh) matrices themselves and in general
infinite-dimensional as is the Hilbert space belonging to the Hamiltonian 
$H_{0}$. 

To give an illustrative example, we consider a two-level system with 
energies $\varepsilon _{\alpha }$ and $\varepsilon _{\beta }$, coupled to a 
time-dependent field via the matrix element $M_{\alpha \beta }=M_{\beta \alpha 
}=2b$, $M_{\alpha \alpha }=M_{\beta \beta }=0$. Taking the limit
$\omega \to 0 $ in Eq.~(\ref{floquet}), we obtain (notice that the quantity
$\Gamma  $ is just a representation of the Green's function $G$ itself):
\begin{eqnarray}
\label{two}
-G^{R}(\omega \to 0) =\left( \begin{array}{ccccccc}
... & & & & & & \\ 
    &\Omega+\varepsilon _{\alpha } &0  &0 &-ib &  &\\
    &0  & \Omega +\varepsilon _{\beta }&-ib  &0 & & \\
    &0 &ib  &\varepsilon _{\alpha } &0  &0  &\\
    &ib &0 &0 &\varepsilon _{\beta } & -ib  &\\
    & & & & & &...  
\end{array}\right)^{-1}. 
\end{eqnarray}
In the case of a coupling proportional to $\cos (\Omega t)$ instead of $\sin 
(\Omega t)$, one easily verifies that $\pm ib$ is replaced by $b$ in 
Eq.~(\ref{two}). One then has exact coincidence with the formal solution of 
the time-dependent two-level problem first given by Shirley~\cite{Shi65} in 
his seminal paper on the Floquet state theory. Here, we have reproduced those 
results in the language of Green's functions starting from the Dyson 
equation. The  Floquet Hamiltonian ${\cal{H}}$ is obtained as the limit 
${\cal{H}}=-(G^{R})^{-1}(\omega \to 0)$ in the representation 
Eq.~(\ref{gamma}) without directly refering to an extended Hilbert space.

\subsection{Dyson equation in second order in the electric field,
electron scattering rate}
We consider Eq.~(\ref{recursion}) to derive the first and second order 
expression for the Green's function $G$.
The first order response to the 
perturbation $M$ is obtained from $g_{0,1}$,
which is an example of how to reproduce results from standard equilibrium 
theory (Kubo's linear response expression) from Keldysh Green's functions.
To next (quadratic) order in $\Sigma $, we obtain the 
$N=0$ (time averaged component) $G(\omega ,N=0)$ from
\begin{eqnarray}
g_{0,0}=g^{0}_{0}-g^{0}_{0}\Sigma [g^{0}_{2}+g^{0}_{-2}]\Sigma g_{0,0},
\end{eqnarray}
which can be written as
\begin{eqnarray}
\label{selfs}
G(\omega ,N=0)&=&\left[(G^{0})^{-1}(\omega )-S(\omega )\right]^{-
1}\nonumber\\
S(\omega ) &:=&\frac{1}{4}WG^{0}(\omega +\Omega )W+(\Omega \to -\Omega ).
\end{eqnarray}
Specializing to the retarded functions, Eq.~(\ref{dysonretarded}), we 
find
$G^{R}(\omega ,N=0)$ $=$ $[(G^{0,R})^{-1}(\omega )-S^{R}(\omega )]^{-
1}$, where $ S^{R}(\omega )$ $:=$ $ (1/4)MG^{0,R}(\omega +\Omega )M+(\Omega \to -
\Omega ) $.

The determination of the Green's function Eq.~(\ref{selfs}) requires a full 
matrix inversion since $S$ in general is a non-diagonal matrix.
The diagonal part of the 
imaginary part of the retarded self energy ${\rm Im}\, S^{R}_{\alpha \alpha 
}$, however,  has adirect physical meaning of an inverse lifetime of the state $\alpha $. It 
describes the scattering rate for scattering out of state $\alpha $ to any 
other eigenstate of the system due to the perturbation. This quantity can be 
expressed exactly by the real part 
$\mbox{\rm Re}\, \sigma _{l}(\varepsilon=\varepsilon_F ,\Omega  )$ of 
the AC conductivity as a function of the Fermi energy $\varepsilon_F$ 
and frequency $\Omega $ {\em in absence} 
of the time-dependent perturbation ${\bf E}_0(t)$. In the limit 
of zero temperature $T=0$, the derivation given in
App. \ref{appendix2} yields 
\begin{eqnarray}
\label{taurelation}
\tau ^{-1}_{MW}=\tau ^{-1}_{+} + \tau ^{-1}_{-}, \quad 
\tau _{\pm}^{-1}(\varepsilon)=
\frac{1}{2}\frac{E_{0}^{2}\mbox{\rm Re}\, \sigma _{l}(\varepsilon\pm 
\hbar\Omega /2,\Omega  )}{(\hbar\Omega )^{2}\rho (\varepsilon 
)}+O\left(\frac{\hbar\Omega }{\varepsilon} \right)^{2}, \quad T=0, 
\end{eqnarray} 
where we reinstalled the $\hbar$ and $\rho (\varepsilon )$ denotes the 
density of states.
This equation relates a physical quantity (the scattering time $\tau _{\pm}$) 
of the non-equilibrium system (static potential + electric field) to a linear 
transport-quantity of the equilibrium system (no electric field). The validity of 
this relation is 
independent of the eigenstate basis and in particular non-perturbative in the 
impurity potential. Basically, the reason why both quantities can be related 
to each other is that both essentially are given by the square of the matrix 
element of the velocity operator. It is also obvious why the 
(Fermi) energy $\varepsilon $ is shifted by $\pm \hbar\Omega /2$ in the 
argument of the conductivity. The latter is due to particle hole excitations 
that lie symmetrically around the Fermi level $\varepsilon $, whereas $\tau 
_{\pm}$ describes electron scattering from $\varepsilon $ to $\varepsilon \pm 
\hbar\Omega$.  Furthermore, the expression for $\tau^{-1}_{MW}$, 
Eq.~(\ref{taurelation}), is valid for a single electron ('empty band'), i.e. 
no Pauli block by the other electrons. If the Fermi sea of the other 
electrons is considered, only scattering through photon absorption ($+$) is 
possible at zero temperature, and $\tau ^{-1}_{MW}=\tau ^{-1}_{+}$. 

The physical meaning of Eq.~(\ref{taurelation}) becomes particularily 
transparent by recalling that the {\em power} $P_{\Omega }$  
absorbed by electrons 
subject to a homogeneous electric field $E_{0}$ is given by
\begin{equation}
\label{powerabsorption}
P_{\Omega }(\varepsilon )=(1/2)E_{0}^{2}{\rm Re}\,\sigma_{l}(\varepsilon )
\cdot L^{d}, 
\end{equation}
where $L^{d}$ is the system volume. For frequencies $\Omega $ 
such that $\hbar\Omega \ll \varepsilon $, 
one can set $\varepsilon \pm \hbar\Omega \approx \varepsilon $, and 
Eq.~(\ref{taurelation}) can be written
\begin{eqnarray}
\label{powerrate}
\frac{P_{\Omega }(\varepsilon )}{\hbar\Omega }=
\frac{\hbar\Omega }{\Delta(\varepsilon ) }\tau _{+}^{-
1}(\varepsilon)
=\rho (\varepsilon ) (L^{d}\hbar\Omega)\, \tau _{+}^{-
1}(\varepsilon),
\end{eqnarray}
where $\Delta(\varepsilon ) :=(L^{d}\rho (\varepsilon ))^{-1}$ is the mean 
level spacing at the energy $\varepsilon $. The number of photons absorbed 
per time, $P_{\Omega }(\varepsilon )/\hbar\Omega $,  is given by the 
scattering rate, multiplied with the density of electronic 
states $\rho (\varepsilon )$ on the scale of the 
photon density of states $(L^{d}\hbar\Omega )^{-1}$. 

\subsection{Microwave radiation in the integer quantum Hall 
effect (IQHE)}
As mentioned in the introduction, in the integer quantum Hall regime, 
microwave induced deviations $\Delta R_{xx}, \Delta R_{xy}$ of the resistance 
tensor have been observed experimentally \cite{Meietal92}. One main 
finding was that these deviations are proportional to the square root of 
the incident microwave power, i.e. linear in the electric field amplitude 
$E_{0}$. 

We calculated the microwave induced scattering rate Eq.~(\ref{taurelation}) 
and found that this linear dependence {\em cannot} be explained in a simple 
scaling picture through the interplay of localization length and inelastic 
length. In fact, in the past the standard scaling picture of the IQHE 
\cite{Pru88} has been widely used to describe the scaling of the temperature 
($T$)--dependent slope $(d\rho _{xy}/dB)^{max}$ and the inverse of the half-
width of the $\rho _{xx}(B)$--peaks \cite{Weietal88,Kocetal91,BSK94}, i.e. 
{\em temperature--dependent} deviations $\Delta R_{xx/xy}$. The main idea is 
that the IQHE as a metal-insulator transition can be described in terms of a 
scaling function, depending on the ratio $\xi/L_{in}$, where $\xi$ is the 
localization length and 
\begin{equation}
\label{lin}
L_{in}=(D\tau )^{1/2}
\end{equation}
is a $T$--dependent inelastic length scale due to electron-electron and/or 
electron-phonon scattering at a rate $\tau^{-1}$. $D$ is a diffusion 
constant. 

Although it is tempting to use this picture under inclusion of the additional 
inelastic scattering process due to the microwave, we found that 
such an explanation definitely is {\em not} appropriate. Our argument is as 
follows. At low temperatures, electron--electron scattering is believed to be 
dominant; in presence of microwave radiation, however, the scattering 
described by the rate Eq.~(\ref{taurelation}) also should contribute. 
If the scattering  due to the microwave is the main
inelastic process, $\tau^{-1}_{MW}\gg \tau _{ee}^{-1}$, where $\tau _{ee}^{-1}$ 
stands for the rate due to  electron--electron, and $\tau ^{-1}_{MW}$ for the 
scattering by the time-dependent field, Eq.~(\ref{taurelation}). One then has
$L_{in}\approx(D\tau _{MW})^{1/2}$ in Eq.~(\ref{lin}), and the scaling 
variable $\xi/L_{in}\sim E_{0}$ becomes proportional to the amplitude $E_{0}$ 
of the microwave field which would hint to the {\em linear} $E_{0}$--
dependence of $\Delta R_{xx/xy}$. On the other hand, the electron-electron 
scattering is dominant for $\tau^{-1}_{ee}\gg \tau _{MW}^{-1}$. 
Estimating the effect by the microwave by adding the scattering {\em rates},
$\tau ^{-1}=\tau _{ee}^{-1}+\tau^{-1}_{MW}$, one finds in this case
\begin{equation}
\label{tau1}
\tau^{-1}=\tau _{ee}^{-1}\left(1+O(E_{0}^{2})\right).
\end{equation}
In particular this would mean that the (small) correction to the scaling variable due 
to the microwave radiation is {\em quadratic} in the amplitude $E_{0}$, and 
so should be any deviation to the resistance $R_{xx/xy}$. 

We used the parameters of the experiment \cite{Meietal92}, 
Eq.~(\ref{taurelation}), and 
a previous calculation of the rate $\tau ^{-1}_{ee}$ \cite{Bra95} to compare 
both quantities. It turns out, that for  temperatures $T\approx 0.5 K$, 
$\Omega = 40 Ghz$, $E_{0}\approx 10^{2} V/m$, the rate $\tau _{MW}^{-1}$ is 
orders of magnitudes smaller than the electron-electron scattering rate $\tau 
_{ee}^{-1}=\gamma k_{B}T/\hbar$ \cite{Bra95}, where $\gamma $ is a factor of 
order unity. Numerically, we find values approximately as $\tau ^{-
1}_{ee}\approx 10^{11} s^{-1}$ and $\tau ^{-1}_{MW}\approx 10^{8} s^{-1}$ for 
an estimated (small) conductivity $\sigma _{xx}\approx 10^{-2}e^{2}/h$. Even 
for a conductivity $\sigma _{xx}$ close to the 'ideal' value $0.5 e^{2}/h$, 
the rate $\tau ^{-1}_{MW}$ stays an order of magnitude behind the electron--
electron scattering rate, and Eq.~(\ref{tau1}) should apply. This, in turn, 
is in contrast to the experimental finding $\Delta R\sim E_{0}$. From this 
negative result we conclude that an interpretation of the experimental data 
\cite{Meietal92} is {\em not} possible in the above picture, simply using 
scaling arguments in combination with inelastic scattering rates. 
We believe that one has to go beyond an explanation that only makes use
of scattering rates, i.e. the diagonal part of the self energy.
Rather, one should start from the full Green's function Eq.~(\ref{floquet})
and investigate directly how quantities like the localization length are
affected by the time dependent perturbation.

\section{Conclusion}
\label{conclusion}
In this paper, we developed a method to calculate the Green's function of 
noninteracting electrons in static potentials in presence of time--dependent 
electric fields. The formalism was given in terms of Keldysh matrices (which 
are block--diagonal for static scattering). Although not discussed here, in 
principal it allows inclusion of interaction processes in form of a suitable 
self energy for electron--electron or electron--phonon interactions. 

In the first approach, by a double Fourier expansion a systematic 
perturbation was developed which is exact in the electric field and sums up 
the static potential to infinite order. Numerical examples for a constant and 
a simple double barrier potential showed that a few Fourier components 
already are sufficient to obtain convergence. We furthermore showed that this 
'truncation method' is a systematic expansion around a non--adiabatic limit 
of high frequencies  $\Omega $ of the electric field. 

A second approach was given in the basis of the exact scattering states. 
There, the truncation method is a perturbation theory in the 
electric field which was explicitly shown for the self energy to second 
order. The latter was used to derive a relation between the rate for
(inelastic) scattering induced by the time dependent field, and the
AC conductivity in absence of the field. Application to microwave experiments 
in the integer Quantum Hall effect suggested, however, that rather than the 
microwave scattering rate the full Green's function in the basis of the 
scattering states has to be investigated. 

Although in both methods exact solutions were derived, the numerical 
applicability suggests that the first approach is appropriate for relatively 
simple static potentials and strong electric fields. The second should be 
used for arbitrary (e.g. random) potentials but rather weak electric fields.  
Although the second approach has not yet been tested numerically, we propose 
to apply it to the localization--delocalization problem. 
 
Discussions with B. Kramer, A. Kawabata, G. Platero, and L. Schweitzer are 
acknowledged. This work has been  supported by the EU STF 9 program and the 
CREST program of the Japan Science and Technology Corporation. 



\begin{appendix}
\section{Derivation of the scattering rate $\tau_{MW}^{-1}$ }
\label{appendix2}
Here, we derive the expression Eq.~(\ref{taurelation}) for the inverse 
lifetime of a state $\alpha $ due to the time dependent electric field.
First, the unperturbed Green's function is
$
G^{0,R}_{\alpha \beta }(\omega )=
\delta _{\alpha \beta }(\omega -\varepsilon _{\alpha }+i0^{+})^{-1}$.
Note that energies are counted from zero and not from a chemical potential
here. The coupling matrix element to the 
electric field is $M_{\alpha \beta }= (E_{0}/\Omega )\langle\alpha|v|\beta 
\rangle$, $v$ is the component of the velocity operator in direction of 
the electric field polarization. This yields
\begin{eqnarray}
\label{imsigma}
\mbox{\rm Im}\, S^{R}_{\alpha \beta }(\omega )&\equiv&
\mbox{\rm Im}\, S^{R+}_{\alpha \beta }(\omega )+
\mbox{\rm Im}\, S^{R-}_{\alpha \beta }(\omega )\nonumber\\
&=&-\pi\frac{E_{0}^{2}}{4\Omega ^{2}}\sum_{\gamma }
\langle\alpha|v|\gamma  \rangle\langle\gamma |v|\beta \rangle
\delta (\omega -\varepsilon  _{\gamma }+\Omega ) +(\Omega \to -\Omega ),
\end{eqnarray}
where we defined the components corresponding to photon absorption
($+ \Omega $ ) and emission $(-\Omega )$
in Eq.~(\ref{imsigma}).

We now recall the expression for the real part of the longitudinal AC 
conductivity in linear response to an arbitrary homogeneous electric field of 
frequency $\Omega $, namely in the zero-temperature limit 
\begin{eqnarray}
\label{linearresponse}
\mbox{\rm Re}\, \sigma _{l}(\varepsilon _{F},\Omega  )&=&
\frac{\pi}{L^{d}}\sum_{\alpha \beta  }
|\langle\alpha|v|\beta \rangle|^{2}
\delta (\varepsilon_{F} +\Omega/2-\varepsilon _{\beta })
\delta (\varepsilon_{F} -\Omega/2-\varepsilon _{\alpha })+O\left(\frac{\Omega 
}{\varepsilon _{F}}\right)^{2}.
\end{eqnarray}
We explicitely indicate the dependence of $\sigma_{l} $ on the Fermi energy,
furthermore $v$ is the component of the velocity operator in one fixed 
direction, say $v_{x}$ for $\sigma _{xx}$. We assume equality of all diagonal 
components of the conductivity tensor and the limit of zero 
temperature; note that the expression explicitely implies symmetry in 
$\Omega $ as it must be since $\mbox{\rm Re}\, \sigma _{l}(\varepsilon 
_{F},\Omega  ) =\mbox{\rm Re}\, \sigma _{l}( \varepsilon _{F},-\Omega  )$. 

The rate for scattering out of a state $\alpha $ with energy $\varepsilon 
=\varepsilon _{\alpha }$ due to photon absorption ($+$) or emission ($-$) 
is given by the diagonal part of the self energy,
\begin{eqnarray}
\label{taudef}
\tau ^{-1}_{MW}=\tau ^{-1}_{+} + \tau ^{-1}_{-}, \quad 
\tau _{\pm}^{-1}(\varepsilon)=-2
\mbox{\rm Im}\, S^{R\pm}_{\alpha \alpha }(\varepsilon  =\varepsilon _{\alpha }).
\end{eqnarray}
The relation between $\tau^{-1}_{MW}$ and the self energy $S^{R}$ is as in the 
equilibrium case and can be directly verified by calculating 
the transition probability from a state $\alpha $ to a state $\beta $ due to 
the time-dependent potential $M(t)$ as in the derivation of Fermi's 'golden 
rule'.

We assume that the diagonal elements of $S^{R}$ depend on the quantum number 
$\alpha $ only through the corresponding eigenenergy $\varepsilon _{\alpha 
}$. In particular, for a (spin-polarized) disorder broadened Landau band 
discussed below, there is a one-to-one correspondence of eigenstates and 
eigenenergies.
From Eq.~(\ref{linearresponse}), one finds
\begin{equation}
\label{compare}
\mbox{\rm Re}\, \sigma _{l}(\varepsilon_{F}\pm\Omega /2,\Omega  )=
\left.\pi\rho(\varepsilon _{F})\sum_{ \beta  }
|\langle\alpha|v|\beta \rangle|^{2}\delta (\varepsilon _{F}\pm\Omega -
\varepsilon _{\beta })\right|_{\varepsilon _{\alpha }=\varepsilon _{F}},
\end{equation}
where $\rho (\varepsilon )=(1/L^{d})\sum_{\alpha }\delta (\varepsilon -
\varepsilon _{\alpha })$ is the density of states.
Combining Eq.~(\ref{imsigma}), Eq.~(\ref{taudef}), and Eq.~(\ref{compare}), 
one obtains Eq.~(\ref{taurelation}).

\end{appendix}

\newpage

\end{document}